\documentclass[aps,prb,twocolumn,superscriptaddress]{revtex4-2}
\usepackage{graphicx}
\usepackage{amsmath}
\usepackage{amssymb}
\usepackage{braket}
\usepackage[colorlinks = true,
            linkcolor = blue,
            urlcolor  = blue,
            citecolor = blue,
            anchorcolor = blue]{hyperref}
\usepackage{comment}
\usepackage{subfigure}
\usepackage[usenames,dvipsnames]{xcolor}
\usepackage{graphicx}
\usepackage{subfigure}
\usepackage{array}
\usepackage[mathscr]{euscript}

\DeclareMathOperator{\Tr}{Tr}

\begin{document}

\title{Slow relaxation of quasi-periodically driven integrable quantum many-body systems}
 
\author{Souradeep Ghosh} 
\email{souradeepg23@tamu.edu}
\affiliation{Department of Physics \& Astronomy, Texas A\&M University, College Station, Texas 77843, USA}
\author{Sourav Bhattacharjee}
\affiliation{ICFO-Institut de Ciencies Fotoniques, The Barcelona Institute of Science and Technology, Av. Carl Friedrich Gauss 3, 08860 Castelldefels (Barcelona), Spain }
\author{Souvik Bandyopadhyay}
\affiliation{Department of Physics, Boston University, Boston, Massachusetts 02215, USA}
%\affiliation{$^4$Department of Physics, Indian Institute of Technology Kanpur, Kanpur 208016, India}

\begin{abstract}
We study the emergence and stability of a prethermal phase in an integrable many-body system subjected to a Fibonacci drive. Despite not being periodic, Fibonacci drives have been shown to introduce dynamical constraints due to their self-similar structure, unlike random driving protocols. From perturbative analysis, this has been argued to result in an exponentially long prethermal phase in the high frequency limit of driving. Examining higher order terms in the perturbative expansion, we show that the perturbative description breaks down eventually in such systems at a finite universal order, which depends solely on features of the Fibonacci sequence. This leads to an onset of energy absorption at long time scales for intermediate and low driving frequencies. Interestingly, in spite of the breakdown of an effective Hamiltonian in the perturbative analysis, we still observe slow logarithmic heating time-scales, unlike purely random drives.

\end{abstract}
\maketitle

\section{Introduction}\label{sec_intro}

%\suvc{May be we should replace "infinite temperature" with "fully mixed" state for all momentum sectors, as we always have translation symmetry?...or do we mention clearly in the introduction what we mean by infinite temperature in rest of the paper?}\\

{With the advent of scalable quantum simulators, there is a new promise of experimentally simulating non-equilibrium quantum many-body systems. This in particular, has resulted in a plethora of recent studies that aim to probe and characterize exclusively out-of-equilibrium phases of matter. One such interesting direction  of research that has gained traction in recent times is the simulation and stabilization of periodically driven many-body systems, within feasible laboratory settings. The rising interest stems from the fact that such systems are well described by the Floquet theory \cite{gomez2013floquet,bukov2015universal,holthaus2016floquet, eckardt2017colloquium, oka2019floquet, rudner2020band,sen21rev,mori2023floquet}, and have been shown to be capable of hosting novel and nontrivial states of matter that are sometimes exclusive to nonequilibrium physics \cite{oka2009photovoltaic, kitagawa2010topological,lindner2011floquet,russomanno2012periodic,thakurathi2013floquet,cayssol2013floquet, bastidas2012nonequilibrium,sharma2014loschmidt,dutta2015statistics,russomanno2015asymptotic,russomanno2016entanglement,sen2016entanglement,gritsev2017integrable,dasgupta2015phase,privitera2016quantum,oka2019floquet, rudner2020band,weitenberg2021tailoring,haldar22review}. \\

The incessant periodic modulation of the Hamiltonian in Floquet systems implies that energy is no longer a strictly conserved quantity in these systems. In generic situations, this often leads to a continuous absorption of energy from the modulation, ultimately driving the system to a featureless  infinite temperature ensemble (ITE) \cite{mukherjee2008defect,d2014long,bhattacharya2018exact,ishii2018heating,lazarides2015fate,haldar2017dynamical,bukov2012parametric}. Exceptions arise in the presence of factors  that can slow down thermalization, such as localization \cite{das2010exotic,d2013many,nag2014dynamical,ponte2015many,haldar21local}, fragmentation, etc \cite{moudgalya2022hilbert,yang2020hilbert,sala2020ergodicity,mukherjee2021minimal,ghosh23frag}. This prethermal phase results from approximately conserved quantities which initially drives the system towards a periodic generalized Gibbs ensemble (pGGE) \cite{rigol2007relaxation,cassidy2011generalized,caux2013time,lazarides2014periodic,nandy18integrable} in the case of integrable systems or a finite temperature Gibbs state of some effective Hamiltonian in the case of nonintegrable systems \cite{d2014long,weidinger2017floquet,haldar18floq,rghosh20perturb}. Importantly, the prethermal phase is found to exist until times which are exponentially long in the driving frequency \cite{else2017prethermal,choi2017observation,weidinger2017floquet,mizuta2019high,kyprianidis2021observation, haldar22review,ghosh23prethermal}.}\\

However in practical situations, it is usually difficult to completely do away with temporal noise in experiments. Imperfect periodicity of the drive may significantly alter the energetics of the system \cite{mukherjee2020restoring}. For example, in the presence of random/completely aperiodic drives, where no local conserved quantity exists, one cannot expect a stable prethermal phase. This, in fact, leads to a very quick build-up of excess energy in the system, ultimately leading to an infinite temperature state \cite{nandy2017aperiodically,maity2018fate,ishii2018heating,dumitrescu2018logarithmically}. The above observation raises the question: how does a system absorb energy when the driving is neither fully periodic nor fully random? Many studies in the last decade in this direction have collectively concluded that there indeed exists a middle ground in the form of quasi-periodic drives, which can also lead to a stable prethermal phase in the system \cite{verdeny2016quasi,nandy2018steady,dumitrescu2018logarithmically,peng2018time,maity2019fibonacci,zhao2019floquet,crowley2019topological,else2020long,lapierre2020fine,boyers2020exploring,crowley2020half,zhao2021random,long2022many}.\\

{Particularly, for a specific type of quasi-periodic modulation-- a binary sequence of two free-fermionic Hamiltonians following a Fibonacci sequence, it was shown in  Ref.~\cite{maity2019fibonacci, nandy2018steady} that at high modulation frequencies, the energy absorbed from the drive saturates to a steady value (until times which are exponentially long in frequency) after an initial transient period. In this paper, we revisit the setup and show that this seemingly steady state, while being certainly long-lived compared to other intrinsic time-scales of the system, does not persist indefinitely and hence can be considered as a prethermal phase. We study the lifetime of the quasi-periodic prethermal plateau and examine how it scales with other relevant time-scales in the problem. Furthermore, we argue that this breakdown of prethermal physics can be attributed to the breakdown of a well-defined quasi-conserved Hamiltonian at sufficiently long times.

Analytically, we find that the prethermal plateau results from a quasi-conserved quantity in the system which has bounded fluctuations. The time-scale associated with this approximate conservation law dictates the onset of heating. Numerical estimation suggests that unlike Floquet systems, heating onsets at a time that scales with the modulating frequency following a power law. 
%In addition, we show that this crossover persists for very large system sizes approaching the thermodynamic limit which we consider to be the largest length scale in our numerical simulations. 
Our investigation further reveals that the existence of the approximate conservation law and the resulting prethermal phase, is an artifact of the self-similar Fibonacci drive itself. In this regard, we note that a similar transition from a quasi-localized to delocalized behavior has also been previously observed for Fibonacci drives in a one-body system, namely a quantum kicked rotor  \cite{bhattacharjee2022quasilocalization} and a disordered many-body chain 
\cite{dumitrescu2018logarithmically} where the authors report logarithmically slow relaxation after the onset of heating. However, our results hint that the quasi-periodic drive induced slow relaxation is a robust phenomenon that can occur even in clean many-body integrable systems. \\

The rest of the paper is organized as follows. In Sec.~\ref{para1}, we define the one-dimensional Kitaev chain and describe the Fibonacci drive protocol briefly. In Sec.~\ref{sec2}, we present the numerical results based on analysing a local observable of the system at various drive frequencies. Sec.~\ref{drive freq}  deals with the dependence of the crossover time (the time at which the system switches from prethermal to growth regime) with the drive frequency and system size. Finally, in Sec.~\ref{sec5} we summarize the results. The appendices at the end recapitulate some previous results relevant to the main text, and also outline some of the calculations involved with more details.

\section{Setup: The Kitaev chain and Fibonacci drive}\label{para1}

We consider a one-dimensional integrable Kitaev chain described by the Hamiltonian \cite{kitaev2001unpaired},
\begin{multline}\label{eq1}
    \mathcal{H}(t) =\displaystyle -\sum_{j=1}^L \left( c_{j}^{\dagger}c_{j+1} + c_{j+1}^{\dagger}c_{j}\right) + \displaystyle\sum_{j=1}^L\Big( c_{j}c_{j+1} \\+ c_{j+1}^{\dagger}c_{j}^{\dagger}\Big)-h(t)\displaystyle\sum_{j=1}^L\Big(2c_j^{\dagger}c_j-1\Big),
\end{multline}
where $c_j$ ($c_j^\dagger$) are the spinless fermionic annihilation (creation) operators, and $h(t)$ is the time-dependent onsite potential. Under periodic boundary conditions, the Hamiltonian in Eq.~\eqref{eq1} decouples in the momentum space into a collection of $N$ non-interacting two-level systems,
\begin{equation*}
\mathcal{H}(t)= \sum_k
    \begin{pmatrix}
    c_{k}^{\dagger} & c_{-k}
    \end{pmatrix}H_k(t)
    \begin{pmatrix}
    c_{k}\\
    c_{-k}^{\dagger}
    \end{pmatrix},
\end{equation*}
such that $k\in[-\pi,\pi]$ .
The single-particle Hamiltonian $H_k(t)$ is of the form,
\begin{equation}\label{eq3}
H_k(t)=(h(t)+\cos k)\sigma^z + (\sin k)\sigma^y,
\end{equation}
where $\sigma_{i}$'s are the Pauli matrices.

\par

In this work, we shall be interested in the stroboscopic dynamics of the system. In other words, we discretize time into intervals of fixed length $T$ and observe the dynamics at stroboscopic instants $t=NT$ for positive integers $N$. The evolution operator $U(N)$ at the stroboscopic instants is thus given as,
\begin{equation*}
U(N)= \hat{\mathcal{T}} \exp{\left[-i \int_{0}^{NT} \mathcal{H}(t) \,dt\right]}\equiv\exp{[-iH(N)NT]},
\end{equation*}
where $\hat{\mathcal{T}}$ is the time-ordering operator and $H(N)$ can be considered as an effective Hamiltonian driving the stroboscopic evolution of the system. For a perfectly periodic drive of the Hamiltonian in Eq.~\eqref{eq1} satisfying $\mathcal{H}(t+T)=\mathcal{H}(t)$, the Kitaev chain attains a periodic steady state $\lim_{N\to\infty}\rho(NT)\equiv \rho_{ss}$. Furthermore, this steady state $\rho_{ss}$ corresponds to a periodic generalised Gibbs ensemble (pGGE) $\rho_{ss}^{GGE}$ \cite{chandran2016interaction,claeys2017breaking,tapias2020out}. This is because the periodicity in the drive guarantees the existence of a time-independent Floquet Hamiltonian $H(N)\equiv H_F$, which satisfies the relation, $\mathcal{T}\exp\left[-i\int_0^T \mathcal{H}(t)dt\right]=\exp\left[-iH_FT\right]$. In addition, the free-fermionic nature of the Hamiltonian guarantees that there exists a time-independent $H_{F,k}$ for each momentum mode, which drives the evolution of the corresponding mode independently when observed at stroboscopic instants, i.e., $\ket{\psi_k(NT)}=\exp[-iNH_{F,k}T]\ket{\psi_k(0)}$. Thus, particle number is conserved and the number of such conserved quantities are extensive in system size $L$. The thermalization to a pGGE is straightforwardly reflected in any local observable $O$, which bears the same underlying decoupled structure in the momentum space as the Hamiltonian,
\begin{multline}\label{obs_def}
%\varepsilon(nT)\equiv (1/L)\sum_k\left[e_k(nT)-e^g_k(0)\right],
\varepsilon_O(NT)\equiv (1/L)\sum_k\Big[\langle\psi_k(NT)|O_{k}|\psi_k(NT)\rangle \\- \langle\psi_k(0)|O_{k}|\psi_k(0)\rangle\Big],
\end{multline}
In the absence of any defect in the periodic drive, $\varepsilon_O$ is found to saturate to a steady value $\lim_{N\to \infty}\varepsilon_O(NT)=\Tr[\rho_{ss}^{GGE}O]$. On the contrary, if periodicity is broken so that no local conserved quantities exist, the system absorbs energy monotonously and approaches an infinite temperature thermal state. However, we note that for finite size systems, the infinite temperature thermal state is never truly achieved due to finite size revivals in the dynamics.  \\

For quasi-periodic drives as in the case of a binary Fibonacci drive, the situation becomes more intriguing. The Fibonacci drive is implemented as follows. We consider a rectangular pulse modulation consisting of two distinct Hamiltonians $H_A$ and $H_B$, the system evolves under either of them in between successive stroboscopic instants. A perfectly periodic drive using such a protocol can be simply implemented as,
\begin{align}
\nonumber	H_{\rm per}(t) &= H_A \quad\quad nT\leq t\leq\left(N+\frac{1}{2}\right)T, \\
		 &= H_B \quad\quad \left(N+\frac{1}{2}\right)T < t \leq (N+1)T.
\end{align} 
Similarly, an aperiodic drive can be emulated by randomly choosing between $H_A$ and $H_B$ for each stroboscopic interval with equal probability. For the quasi-periodic drive, the sequence of $H_A$ and $H_B$ follows a binary Fibonacci sequence (see Appendix~\ref{appA}). To elaborate, the driving sequence upto any given $t=NT$, where $N=\mathcal{F}_\mathcal{N}\in\{1,2,3,5,8,13,\cdots\}$, is given by the sequence $S_\mathcal{N}$, which satisfies the Fibonacci recursion relation $S_{\mathcal{N}+1}=S_{\mathcal{N}}S_{\mathcal{N}-1}$ as demonstrated below, 
\begin{align}
	 S_{\mathcal{N}=1}&=H_A\nonumber\\
	  S_{\mathcal{N}=2}&=H_AH_B\nonumber\\
	   S_{\mathcal{N}=3}&=H_AH_BH_A\nonumber\\
	    S_{\mathcal{N}=4}&=H_AH_BH_AH_AH_B\nonumber\\
	     S_{\mathcal{N}=5}&=H_AH_BH_AH_AH_BH_AH_BH_A\nonumber\\
 S_{\mathcal{N}=6}&=H_AH_BH_AH_AH_BH_AH_BH_AH_AH_BH_AH_AH_B,
\end{align} 

As in the case of random drives, it is not possible to identify any time-independent effective Hamiltonian, at least stroboscopically, dictating the dynamics of the system. Nevertheless, it was previously reported in Ref.~\citep{maity2019fibonacci,nandy2018steady} that in the limit of high driving frequencies $T\to 0$, one can identify an effective conserved Hamiltonian (with fluctuations) that bounds the local observable $\varepsilon_O(NT)$ within a finite range of values. However, as we shall demonstrate in the following sections, this remains true only when the dynamics can be approximated by a Baker-Campbell-Hausdorff expansion of the evolution operator $U(N)$ \cite{klarsfeld1989baker} up to linear order in stroboscopic interval $T$. Furthermore, even in the high frequency limit, this observable does not remain bounded forever and eventually heats up to infinite temperature.  It is to be noted that by infinite temperature state, we mean, the density matrix behaves like a maximally mixed state in each $k$-sector, which is the maximum entropic state in this situation. In fact, this is a generic feature of the Fibonacci drive and similar results have been reported in the dynamics of a quantum rotor kicked with Fibonacci sequence~\citep{bhattacharjee2022quasilocalization}; where the kinetic energy of the rotor was found to be initially localised at high drive frequencies before an eventual growth is observed. 

\begin{figure}[t]
	
	\includegraphics[width=0.5\textwidth]{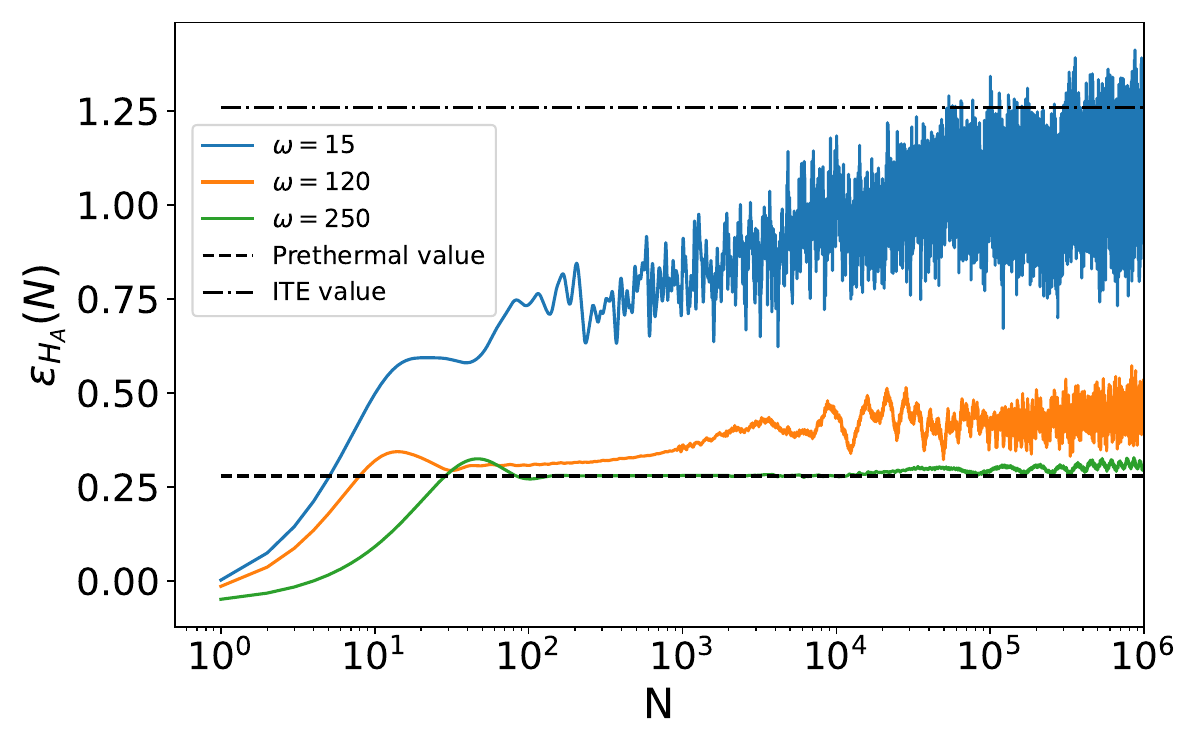}
	
	\caption{Local observable $\varepsilon_{H_{A}}(N)$  plotted for various frequencies as a function of N with $h_A=1, h_B=10, L=100$ and the initial state taken to be the ground state of Hamiltonian $H_A$. The length of the pre-thermal region increases as frequency is increased. On the other hand, the observable $\varepsilon_{H_A}(N)$ approaches its infinite temperature value at long times. Note that the rate of energy growth gets slower as we increase the driving frequency. The noise due to the finite size effects has been filtered out for better visual understanding, using the pre-defined filtfilt() function of Scipy Signal library in Python.}\label{RE_compare}
\end{figure}

\section{The fate of the local observable $\varepsilon_O(N)$}\label{sec2}

In this section, we shall demonstrate that the non-equilibrium steady state (NESS) reported previously in Ref.~~\citep{maity2019fibonacci,nandy2018steady} is in fact a prethermal state that eventually heats up to infinite temperature. We shall supplement the numerical findings with detailed analysis of progressively higher order BCH expansions of the evolution operator $U(N)$. In particular, we shall be interested in the evolution operator for any generic mode $U_k(N)$, which is given by,
\begin{multline}\label{eq_ham_effective}
    U_k(N)=\exp\Big[-iH^A_kT-iH^B_kT\\-i\left[H^A_{k},H^B_{k}\right]T^2-\dots\Big]\equiv\exp\left[-iH^{Fib}_k(N)NT\right]
\end{multline}

\subsection{Prethermal regime in the high frequency limit}\label{high freq}

Let us first consider the case when $T\to 0$ ($\omega=T^{-1} \gg 1$). For numerical simplicity, we choose the local observable $O$ in $\varepsilon_O(N)$ as the Hamiltonian $H_A$ itself with $O_k=H^A_{k}$ (see Eq.~\eqref{obs_def}). Without loss of generality, we also choose the initial states of each momentum mode to be the ground state of Hamiltonian $H^A_{k}$, with system size $L=100$ and $h_A=1$; $h_B=10$. As shown in Fig.~\ref{RE_compare}, $\varepsilon_{H_A}(NT)$ initially tends towards a prethermal value (shown by the dotted lines) with bounded fluctuations (the steps to obtain this prethermal value analytically can be found in Ref.~\cite{maity2019fibonacci} or in Appendix \ref{appC}). However, it eventually starts to grow unboundedly ; the initial saturation thus suggests that the system goes through a pre-thermal regime before approaching an infinite temperature state in each momentum block. The infinite temperature value can be evaluated by modifying Eq.~\eqref{obs_def}:

\begin{eqnarray}
\nonumber \varepsilon_{H_A}(N \to \infty)=\frac{1}{L}\sum_k \Big[ \Tr[H^A_k]- \langle \nonumber\psi_k(0)|H^A_k|\psi_k(0) \rangle \Big] \\
= -\frac{1}{L}\sum_k\langle \psi_k(0)|H^A_k|\psi_k(0) \rangle,
\end{eqnarray}

However, we see that reaching this infinite temperature ergodic state can be exponentially slow for the Fibonacci drive. For high driving frequencies, the time taken to reach this steady state is beyond our simulation times for large system sizes.  It is expected that a qualitatively similar behaviour would be observed if one identifies the local observable as  $H_B$, such that $O_k=H^B_{k}$. The pre-thermal regime arises due to the existence of pseudo-conserved quantities during the initial phase of the evolution \citep{maity2019fibonacci}. To illustrate the same, let us consider a BCH expansion up to order $\mathcal{O}(T^2)$ of the evolution operator, and extract the Hamiltonian $H_k^{Fib}(N)$ (up to order $\mathcal{O}(T)$) in Eq.~\eqref{eq_ham_effective} as, 
%\begin{multline}\label{eq4}
%    U_k(N) \approx \exp\{ -iT\alpha(N)H_k^A - iT\beta(N)H_k^B\\ - T^2 \delta(N)[H_k^A,H_k^B]+\mathcal{O}(T^3) \} 
%\end{multline}
%We can rewrite Eq.~\eqref{eq4} as $U_k(N)\equiv \exp{\bigl[-iNTH_{k}(N)]}$, where $H_k(N)$ is the effective Hamiltonian at the $N$-th stroboscopic interval, given by :
\begin{equation}\label{eq_ham_first}
    H_k^{Fib}(N) \approx \frac{\alpha(N)}{N}H_k^A + \frac{\beta(N)}{N}H_k^B - \frac{iT\delta(N)}{N}[H_k^A,H_k^B].
\end{equation}
In the above equation, the coefficients $\alpha(N)$, $\beta(N)$ and $\delta(N)$  \cite{maity2019fibonacci} satisfy,
\begin{align}
\alpha(N)&=\displaystyle\sum_{m=1}^N \{\gamma(m)-1\},\\
\beta(N)&=\displaystyle\sum_{m=1}^N \{2-\gamma(m)\}=N-\alpha(N),\\
\delta(N)&=\frac{1}{2}\displaystyle\sum_{m=1}^N\{(\gamma(m)-1)(m-1)-\bigl \lfloor \frac{mG}{1+G}\bigr \rfloor\},
\end{align}

\begin{figure*}[t]
	\centering
	\subfigure[]{
		\includegraphics[width=0.44\textwidth]{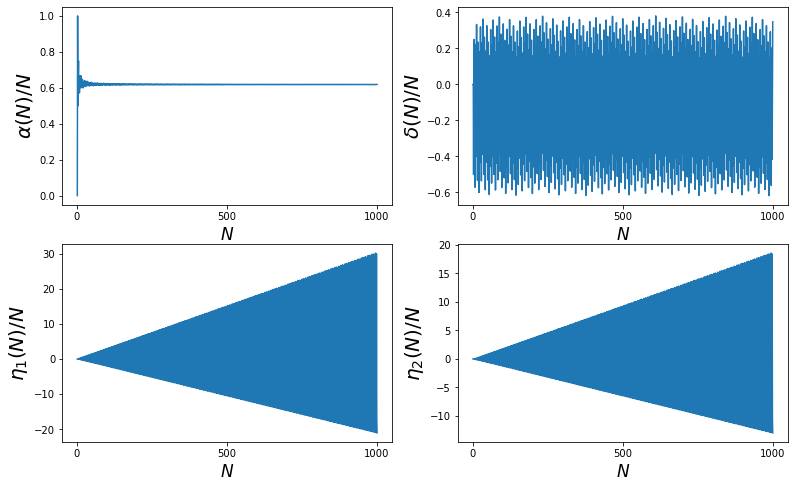}
		\label{NEC plot}}
	\subfigure[]{
		\includegraphics[width=0.44\textwidth]{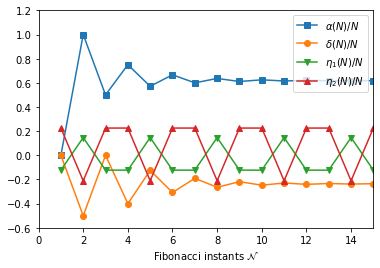}\label{NEC fib plot}}
	\caption{(a) The Normalised Expansion Coefficients behaviour with time (at stroboscopic intervals N). $\alpha(N)/N$ saturates to a constant value with time, $\delta(N)/N$ oscillates between two values but stays bounded, and the other two grow unboundedly with time. (b) The NECs observed at Fibonacci instants $\mathcal{N}$. }
\end{figure*}

where, $\gamma(m)=\bigl \lfloor (m+1)G\bigr \rfloor-\bigl \lfloor mG\bigr \rfloor$, and $G$ is the Golden ratio (The ratio of consecutive Fibonacci numbers approaches $G=\frac{1+\sqrt{5}}{2}$ at large $N$). It is evident from Fig.~\ref{NEC plot} that  $\alpha(N)/N$ (and consequently $\beta(N)/N$) saturate to constant values after a very short time. On the contrary, $\delta(N)/N$ fluctuates rapidly between $(1-1/G)\approx 0.382$ and $-1/G \approx -0.618$. These observations suggest that as long as the Hamiltonian in Eq.~\eqref{eq_ham_first} approximates the full effective Hamiltonian $H_k(N)$, the evolution is driven by a time-independent Hamiltonian (with small fluctuations). The observable $\varepsilon_O(N)$ in this case is expected to saturate with fluctuations about a mean value. Indeed, at a moderately large $N$, the contribution of $\delta(N)/N$ vanishes upon summation over the $k$-modes, and the system reaches a steady state value \citep{maity2019fibonacci}; a brief outline of the analytical steps involved in calculating these Normalised Expansion Coefficients (NECs) is provided in Appendix~\ref{appC}.\\ 

 It is also important to note that there are two crucial conditions that are needed to be satisfied in order for the first order expansion of the effective Hamiltonian in Eq.\eqref{eq_ham_first} to approximate the full Hamiltonian in Eq.~\eqref{eq_ham_effective}. Firstly, the driving frequency must be high, i.e., $T\to 0$ and secondly, the normalized coefficients of the (ignored) higher order terms must not diverge with $N$. In the following, we shall demonstrate that it is the precisely the failure of the latter condition which leads to a breakdown of the prethermal regime.

\subsection{Breakdown of the BCH expansion at low frequencies}\label{sec low freq}
 From Fig.~\ref{RE_compare}, it is evident that the prethermal regime gives way to a regime of unbounded growth of the observable $\varepsilon_{H_A}(N)$ at sufficiently long times; with the crossover time decreasing as the driving frequency is lowered. 
To explain this, let us analyze the higher order terms in the BCH expansion of the evolution operator $U_k(N)$. To begin with, we consider an expansion up to order $\mathcal{O}(T^3)$ of the evolution operator (order $\mathcal{O}(T^2)$ in $H^{Fib}_k(N)$),
\begin{multline}\label{eff h1}
    H^{Fib}_k(N) \approx \frac{\alpha(N)}{N}H_k^A + \frac{\beta(N)}{N}H_k^B - \frac{iT\delta(N)}{N}[H_k^A,H_k^B] \\- \frac{T^2\eta_1(N)}{N}[H_k^A,[H_k^A,H_k^B]] - \frac{T^2\eta_2(N)}{N}[H_k^B,[H_k^B,H_k^A]+\mathcal{O}(T^3),
\end{multline}
Here, the coefficients $\eta_1(N)$ and $\eta_2(N)$ can be evaluated as \cite{bhattacharjee2022quasilocalization} (see Appendix \ref{appB} for a derivation) :

\begin{widetext}
\begin{subequations}
\begin{equation}
\eta_1(N)=\frac{1}{12}\displaystyle\sum_{m=1}^{N}\bigl[(2-\gamma(m))\bigl \lfloor \frac{mG}{1+G}\bigr \rfloor^2 - (1-\gamma(m))\bigl(6\delta(m-1)+(2-m)\bigl \lfloor \frac{m}{1+G}\bigr \rfloor + \bigl \lfloor \frac{m}{1+G}\bigr \rfloor^2 \bigr) ]\label{eta1}
\end{equation}
\begin{equation}
\eta_2(N)=\frac{1}{12}\displaystyle\sum_{m=1}^{N}\bigl[(\gamma(m)-1)\bigl \lfloor \frac{m}{1+G}\bigr \rfloor^2 + (2-\gamma(m))\bigl(-6\delta(m-1)+(2-m)\bigl \lfloor \frac{mG}{1+G}\bigr \rfloor + \bigl \lfloor \frac{mG}{1+G}\bigr \rfloor^2 \bigr) ] \label{eta2}
\end{equation}
\end{subequations}
\end{widetext}

From Fig.~\ref{NEC plot}, it appears that that these NECs, $\eta_1(N)/N$ and $\eta_2(N)/N$ grow unboundedly with $N$, thereby ruling out the possibility of approximating local dynamics by an $N$-independent Hamiltonian. However, if one observes the NECs at Fibonacci instants $F \mathcal{(N)}$ in stead of stroboscopic instants $N$, they are found to remain bounded. This is seen in  Fig.~\ref{NEC fib plot}, where $\beta(F(\mathcal{N}))/F(\mathcal{N})$ and $\delta(F(\mathcal{N}))/F(\mathcal{N})$ saturate to a steady value after $\mathcal{N} > 10$, and the coefficients $\eta_1(F(\mathcal{N}))/F(\mathcal{N})$ and $\eta_2(F(\mathcal{N}))/F(\mathcal{N})$ oscillate between two constant values. 

This implies that one will find the observable $\varepsilon_{O}(N)$ to remain saturated when observed at Fibonacci instants even though the time interval between successive Fibonacci instants grow exponentially at long times. Furthermore, by construction the sequence of $H_A$ and $H_B$ in between two successive Fibonacci instants $F(\mathcal{N})$ and $F(\mathcal{N}+1)$ is also a Fibonacci sequence itself, namely the sequence $S_{\mathcal{N}-1}$. Thus, considering the state of the system at the stroboscopic instant $N=F(\mathcal{N})$ as the initial state, $\varepsilon_O(N)$ will also be found to be saturated at $N=F(\mathcal{N})+F(\mathcal{N}'=1), F(\mathcal{N})+F(\mathcal{N}'=2), F(\mathcal{N})+F(\mathcal{N}'+3), \dots$, where $
\mathcal{N}'$ denotes the Fibonacci instants nested between $F(\mathcal{N})$ and $F(\mathcal{N}+1)$. Similarly, the sequence between  $F(\mathcal{N}')$ $F(\mathcal{N}'+1)$ is also a Fibonacci sequence and using the same argument as above, one can argue that the local observable $\varepsilon_O(N)$ will also be saturated at Fibonacci instants nested between these two instants. Therefore, the self-similar structure of the Fibonacci drive ensures that the observable $\varepsilon_O(N)$ will be saturated not only at Fibonacci instants $F(\mathcal{N}=1)$ but also effectively at all stroboscopic instants. In fact, one can estimate that this fractal nature of the Fibonacci drive results in an exponentially stable prethermal phase in the high frequency limit (see Ref.~\cite{maity2019fibonacci}) if one considers the effective Hamiltonian expansion upto order ${\rm O}(T^2)$.\\

\begin{figure*}[t]\label{bloch}
	\centering
	\subfigure[]{
		\includegraphics[width=0.25\textwidth]{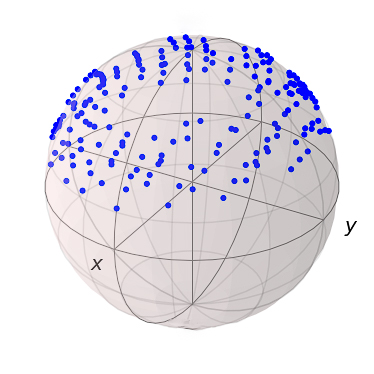}
		\label{blocha}}	
	\centering
	\hfill
	\subfigure[]{
		\includegraphics[width=0.25\textwidth]{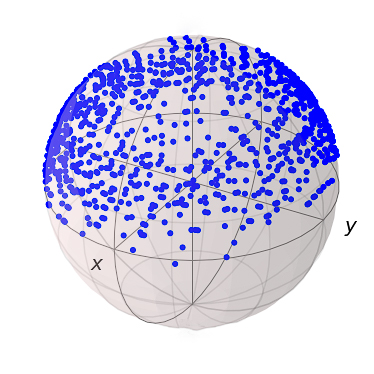}
		\label{blochb}}
	\centering
	\hfill
	\subfigure[]{
		\includegraphics[width=0.25\textwidth]{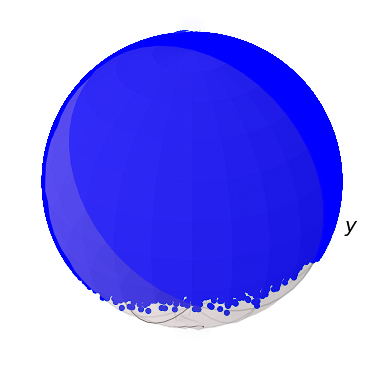}
		\label{blochc}}	
	\caption{Trajectory of $|\psi_k(N)\rangle$, with $k=159\pi/200$, at $\omega=55$ after (a) $N=200$, (b) $N=1000$ (c) $N=10^6$ stroboscopic intervals (here, the initial state is taken to be the ground state of Hamiltonian $H_A$, and system size $L=100$ ). For a given intermediate frequency, the trajectory is bounded between two concentric circles initially. As the time of observation is increased, the area of this annular region increases, until eventually the trajectory starts covering the entire Bloch sphere uniformly. For higher driving frequencies, the deviation from the annular region occurs even at larger times, signaling very slow relaxation to the infinite temperature ergodic state.} %Note that a similar transition would be observed if the number of observed stroboscopic intervals is kept fixed and the driving frequency is decreased.}
\end{figure*}

Likewise, we consider the new terms appearing from an expansion of the effective Hamiltonian to order $\mathcal{O}(T^3)$. The NECs corresponding to the commutators $[H^A,[H^A,[H^A,H^B]]]$, $[H^B,[H^B,[H^A,H^B]]]$ and $[H^A,[H^B,[H^A,H^B]]]$  grow unboundedly even when observed at Fibonacci instants  $F(\mathcal{N})$; analytically, the respective NECs are given by \cite{dumitrescu2018logarithmically},
\begin{widetext}
\begin{subequations}\label{mu}
\begin{equation}
\frac{\mu_1(F(\mathcal{N}))}{F(\mathcal{N})}=\frac{(-1)^{\mathcal{N}}}{120}\left[G^{\mathcal{N}-1}+\frac{1}{G}[(-1)^{\mathcal{N}}(3G-4)-1-3G] \right],\label{mu1}
\end{equation}
\begin{equation}
\frac{\mu_2(F(\mathcal{N}))}{F(\mathcal{N})}=\frac{(-1)^{\mathcal{N}}}{120}\left[G^{\mathcal{N}-1}(2-G)+\frac{1}{G}[(-1)^{\mathcal{N}}(4G-7)-2-G] \right],\label{mu2}
\end{equation}
\begin{equation}
\frac{\mu_3(F(\mathcal{N}))}{F(\mathcal{N})}=\frac{(-1)^{\mathcal{N}}}{120}\left[2G^{\mathcal{N}-1}(1-G)+\frac{1}{G}[(-1)^{\mathcal{N}}(3-G)+3+4G] \right],\label{mu3}
\end{equation}
\end{subequations}
\end{widetext}

where we have ignored terms of order $G^{-(\mathcal{N}+1)}, G^{-(2\mathcal{N}+1)}$,etc. The leading order term in these NECs grow as $G^{\mathcal{N}}$, i.e., they are unbounded.

The breakdown of the prethermal regime can now be explained as follows. In the prethermal regime, the contribution from the terms of order $\mathcal{O}(T^3)$ (and above) is negligible as they are suppressed by a factor of $T^3$ and higher powers of $T$ for sufficiently high driving frequencies. However, as the NECs themselves grow in magnitude with $N$, these terms eventually begin to contribute significantly and hence the Hamiltonian expanded upto order $\mathcal{O}(T^2)$ no longer approximates the full effective Hamiltonian. As a result, approximate conservation laws are destroyed and can no longer constrain local dynamics.  Hence, it is evident that when the effect of these higher order NECs in the BCH expansion (Eq.~\eqref{eff h1}) comes into play, the system crosses over from the prethermal regime to a regime where it starts to absorb energy from the drive. This explains the observation in Fig.~\ref{RE_compare} that the stability of the pre-thermal region decreases as the driving frequency is lowered. \\

\par This transition from pre-thermal to {heating} regime can also be understood if we represent the time-evolution unitary operator as a Bloch vector, $U_k(N)=e^{i\theta_{N,k}\hat{e}_{N,k}\cdot\vec{\sigma}},$ where $\theta_{N,k}$ and $\hat{e}_{N,k}$ changes with stroboscopic intervals $N$, and $\sigma$'s are the Pauli matrices. If we write the initial ground state of the system (here, taken to be of the Hamiltonian $H^A_{k}$ with $h_A=1$, $h_B=1$) in the form of :
\begin{equation*}
|\psi_k(0)\rangle=
\begin{bmatrix}
    \cos(\theta_k(0)/2)\\
    \sin(\theta_k(0)/2)e^{i\phi_k(0)}
\end{bmatrix},
\end{equation*}
which denotes a point on the Bloch sphere with coordinates $[x_0,y_0,z_0]\equiv[\sin(\theta_k(0))\cos(\phi_k(0)),\sin(\theta_k(0))\sin(\phi_k(0)),\cos(\theta_k(0))]$. Evolving this initial state with the unitary operator $U_k(N)$, we can track the trajectory of the state $|\psi_k(N)\rangle$ on the Bloch sphere (or, $[x_t,y_t,z_t]$), as it evolves in time (for a single $k$-mode). On doing so, we see from Fig.~\ref{blocha}, that for a moderately high frequency, the trajectory oscillates within an area bounded by two concentric circles, which is due to the oscillatory nature of NEC $\delta(N)/N$. In other words, the trajectory remains bounded as the quasiperiodic drive emulates a periodic drive in the high frequency limit and is therefore constrained by the existence of the pseudo-conserved quantities \cite{maity2019fibonacci}. However, as we wait for a longer time, we see that the annular region starts broadening (Fig.~\ref{blochb}) and eventually the trajectory covers almost the entire Bloch sphere uniformly (Fig.~\ref{blochc}), implying the breakdown of the prethermal region and the delocalization of the state vector over the complete Hilbert space.  Therefore, the system eventually starts absorbing energy and under generic settings Fibonacci circuits lead to complete mixing in Hilbert space \cite{choi23}. However, we find that relaxation to this infinite temperature state can be  very slow, unlike random processes.\\

\par The local observable $\varepsilon_{H_A}(N)$ is difficult to measure experimentally. We therefore supplement the above results with a numerical analysis of the particle number density (with an offset) of the system:
\begin{equation}\label{mag_real}
\langle m(N) \rangle = \frac{1}{L}\sum_i \langle\psi(N)|\left(c_i^{\dagger}c_i-\frac{1}{2}\right)|\psi(N) \rangle ,  
\end{equation} 
We can rewrite the corresponding expression in the momentum space as,
\begin{equation}\label{mag_k}
\langle m(N) \rangle = \frac{1}{L}\sum_k \langle \psi(N)|\left(c_k^{\dagger}c_k-\frac{1}{2}\right)|\psi(N) \rangle
\end{equation}
Starting from the initial state to be a product state $ \ket{ 1111...}$, we evolve the system quasi-periodically with the Fibonacci kicks. When the particle number density is plotted against time (or stroboscopic intervals) $N$ (Fig.~\ref{mag_compare}), it is observed that after saturating to an initial prethermal value, number density starts to deviate from it's prethermal value and approaches that of the maximum entropic completely mixed state for every momentum sector. However, interestingly, the approach towards zero (which is characteristic of an infinite temperature steady state) is very slow (consistent with $\log N$) in contrast to a purely random drive.  
\begin{comment}
This is because of the fact that when the system mimics the behaviour of a random drive, its density matrix turns into a unit matrix. Since $\sigma^z$ is traceless, its expectation value is zero, i.e., $\langle \sigma^z \rangle = \text{Tr}[\rho\sigma_z]=0$. 
\end{comment}
Also, we note that the duration of the prethermal region of the number density increases as the drive frequency is increased. On the contrary, for a low driving frequency, no prethermal saturation is visible at all numerically.

\begin{figure}[t]
	\centering
	
	\includegraphics[width=0.45\textwidth]{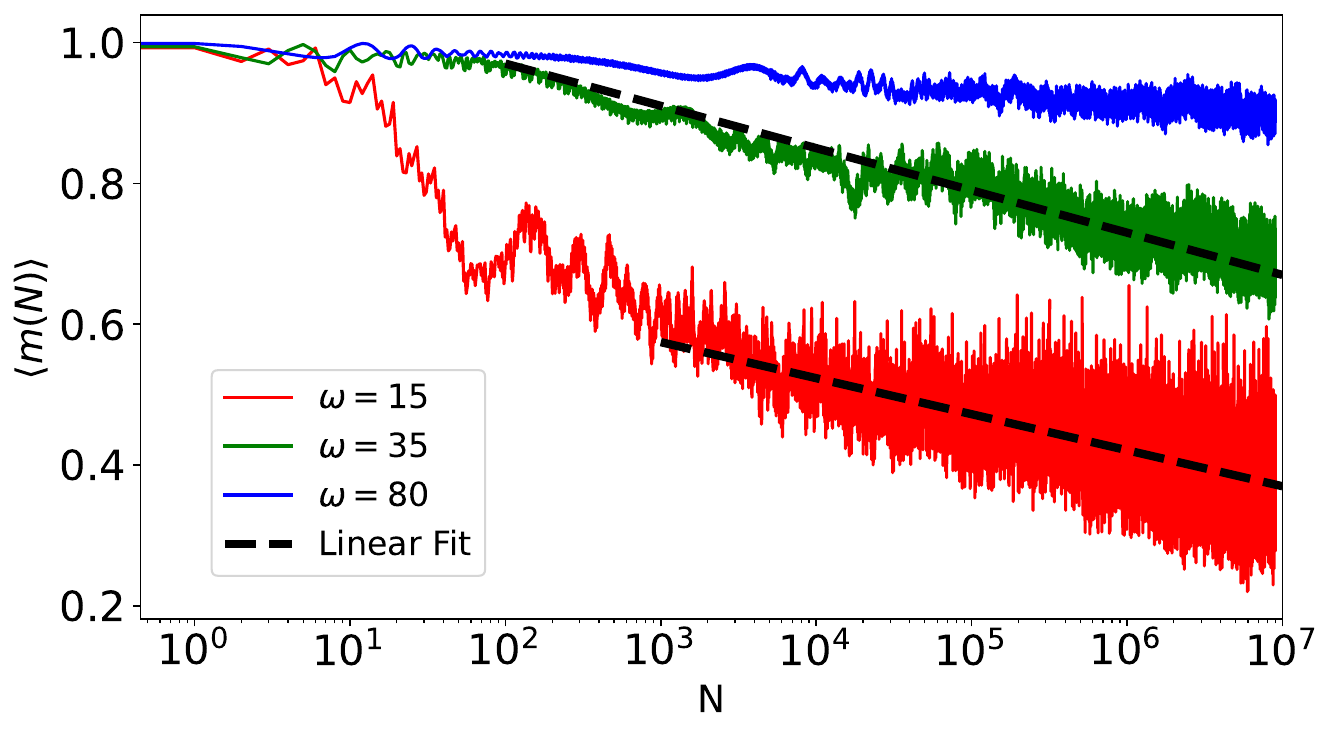}
		
	\caption{Offset particle number density $\langle m(N) \rangle$ plotted as a function of time (or stroboscopic intervals) $N$, with the initial state as a product state $ \ket{ 1111...}$, system size $L=400$, and $h_A=1, h_B=10$. We see that the particle number density starts to dip to a near zero value when the prethermal regime breaks down. Upto the region of our observation ( $N=10^7$ intervals), as shown by the linear fit, $\langle m(N) \rangle$ exhibits a slow logarithmic relaxation with N. }

	\label{mag_compare}
\end{figure}

\section{Variation of the cross-over time with frequency}\label{drive freq}

\begin{figure}[t]
	\centering
	\includegraphics[width=0.48\textwidth]{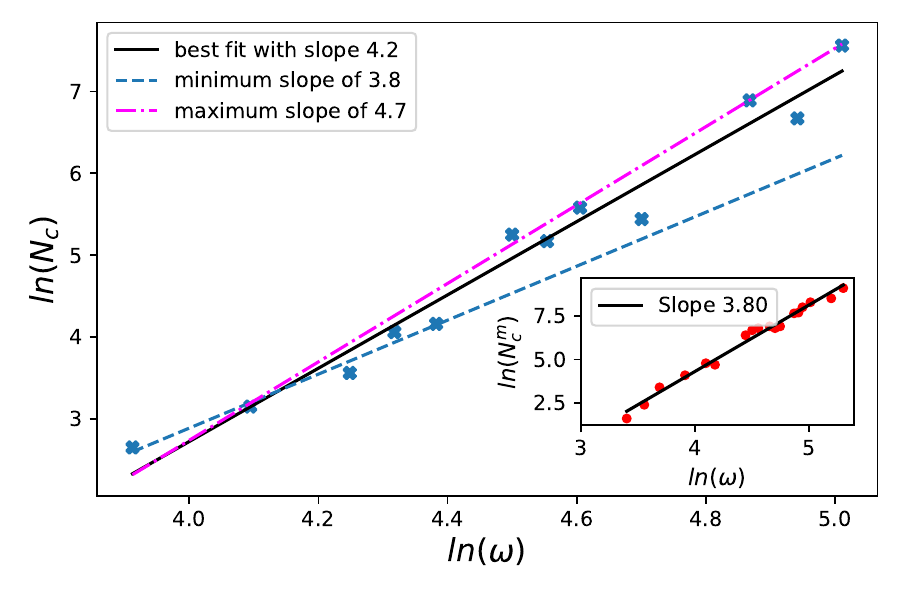}
	
	\caption{Crossover time $N_{c}$ as a function of frequency $\omega$, with $L=100$. The above straight line (in the log-log scale) has slope $=4.2\pm 0.9$, which implies a power law relation $( N_{c} \propto \omega^{4.2 \pm 0.9} )$; the inset shows the scaling of cross-over time of particle number density $N_c^m$ as a function of frequency to be a straight line on a log-log scale, which again implies a power law relation $N_c^m \propto \omega^{3.8}$.}
	\label{fig_freqb}
\end{figure}

We have shown in Sec.~\ref{sec low freq}, that the saturated or bounded values of NECs at Fibonacci instants $\mathcal{N}\gg1$, is necessary for a steady state behaviour and the existence of an effective Fibonacci Hamiltonian. The first five NECs (upto the order $\mathcal{O}(T^2)$), are either saturated or bounded between two constant values when observed at Fibonacci instants $F(\mathcal{N})$. However, the $\mathcal{O}(T^2)$ grow unboundedly when observed at $N=F(\mathcal{N})$. Without explicitly evaluating the commutators, we can make an order of magnitude estimate of the approximate time after which the system switches over from prethermal to growth regime. Let us dub this cross-over time to be $N_{c}$. From Eq.~\eqref{mu}, we can see that leading-order unbounded term in the expansion coefficient of $\mathcal{O}(T^3)$, is of the order $G^{\mathcal{N}}$. Again, the stroboscopic instant $N$ corresponding to Fibonacci instant $\mathcal{N}$ is given as $N\approx G^{\mathcal{N}}$ (see Appendix ~\ref{appA}). Hence, the $\mathcal{O}(T^3)$ terms become significant when $T^3N_{c}\sim1$, so $N_{c}\propto\omega^3$, i.e., an increasing {power law} dependence with drive frequency.
\par Numerically, we can find a rough estimate of $N_{c}$ for a given $\omega$ by fitting a straight line into the growth regime of the local observable $\varepsilon_O(N)$ obtained by exact diagonalization of the Hamiltonian, and finding out its point of intersection with the $\varepsilon_O(N)$ obtained with only the bounded terms of the Hamiltonian (see Appendix~\ref{appD}). Similarly, we find the $N_{c}$ for other frequencies, and plot them on a log-log scale. As shown in Fig.~\ref{fig_freqb}, we get a straight line with slope $=4.2 \pm 0.9$, i.e., $N_{c} \propto \omega^{4.2 \pm 0.9}$ implying a {power law} dependence as predicted theoretically (see Appendix ~\ref{appE} for a discussion on the dependence of prethermal lifetime on the commutator norm of the effective Hamiltonian). A similar analysis of the particle number density shows that the time $N_c^m$ at which $\langle m \rangle$ starts to dip varies with the driving frequency as $N_c^m\propto \omega^{3.8}$ (see inset of Fig.~\ref{fig_freqb}). \\

\par The slight mismatch between the theoretically predicted value of the power law exponent and that extracted numerically is due to the fact that the crossover between pre-thermal and growth regime is not sharp and is therefore difficult to pinpoint. Moreover, the noise arising from finite size effects also adds a scope of uncertainty. Additionally, the theoretical reasoning is but a perturbative analysis, and it is only possible to determine the minimum order of terms in the BCH expansion before the system shows heating behaviour. In our case, we can only say that the terms of order equal to or higher than $\mathcal{O}(T^3)$ are responsible for the onset of the growth regime.\\

\section{Concluding comments}\label{sec5}
In summary, we probe the stability of a prethermal phase in Trotterized Fibonacci driving protocols in an integrable system. Specifically, we observe the dynamics of local observables through their expectation values in a free fermionic model under periodic boundary conditions, when subjected to a Fibonacci drive. The main observation of the analysis is that unlike random driving, even though the self-similar structure of a Fibonacci sequence induces a prethermal phase in local observables, the system eventually heats up towards a maximal entropic state in each momentum sector. This is primarily because, one can usually describe the dynamics of such systems using a high frequency expansion to obtain a quasi-conserved local Hamiltonian, which in turn dictates prethermal physics. However, we show that any notion of a  quasi-conserved effective Hamiltonian must break down after a finite time in such systems. Furthermore, this breakdown of the effective Hamiltonian is solely a universal feature of the Fibonacci sequence and are therefore independent of the driving Hamiltonians altogether. Remarkably, we find that the relaxation from the prethermal phase to the infinite temperature state is logarithmically slow in time.\\
	
Interestingly, we numerically establish that the time after which the system starts absorbing heat goes as a power-law in frequency. Furthermore, we believe that this breakdown of an effective Hamiltonian is a universal property of the Fibonacci sequence and does not depend on details of the system studied. We also demonstrate this breakdown of prethermal physics through delocalization of the time-evolved state over the Hilbert space (constrained within each momentum sector) and also the asymptotic steady state of fermion occupations.\\
	
The emergent universality in the breakdown of the high frequency effective Hamiltonian expansion suggests that it would be interesting to study similar dynamics in more generic nonintegrable systems. Another direction of future research is to explore if it is possible to systematically find driving protocols that can be classified to have similar universal breakdown of an effective hamiltonian picture. We further note that given state-of the art quantum simulators, a simulation of the complete dynamics is very much accessible to current day experiments \cite{zhao2022making,zhao2023making,barbiero2020bose}.\\

\acknowledgements

We reminisce with love the memory of Amit Dutta with whom the discussions on this project
started. We thank Utso Bhattacharya, Somnath Maity, Diptiman Sen, Adhip Agarwalla, Diptarka Das, Saikat Mondal, Anatoli Polkovnikov, Anushya Chandran, Joaquin F. Rodriguez-Nieva and Christopher Laumann for discussions and fruitful comments.. S. Ghosh acknowledges Texas A\&M University for financial support and Grace cluster of TAMU HPRC where parts of the numerical simulations were carried out. S. Bhattacharjee acknowledges support from: European Research Council AdG NOQIA; MCIN/AEI (PGC2018-0910.13039/501100011033, CEX2019-000910-S/10.13039/501100011033, Plan National FIDEUA PID2019-106901GB-I00, Plan National STAMEENA PID2022-139099NB, I00,project funded by MCIN/AEI/10.13039/501100011033 and by the “European Union NextGenerationEU/PRTR" (PRTR-C17.I1), FPI); QUANTERA MAQS PCI2019-111828-2); QUANTERA DYNAMITE PCI2022-132919, QuantERA II Programme co-funded by European Union’s Horizon 2020 program under Grant Agreement No 101017733); Ministry for Digital Transformation and of Civil Service of the Spanish Government through the QUANTUM ENIA project call - Quantum Spain project, and by the European Union through the Recovery, Transformation and Resilience Plan - NextGenerationEU within the framework of the Digital Spain 2026 Agenda; Fundació Cellex; Fundació Mir-Puig; Generalitat de Catalunya (European Social Fund FEDER and CERCA program, AGAUR Grant No. 2021 SGR 01452, QuantumCAT \ U16-011424, co-funded by ERDF Operational Program of Catalonia 2014-2020); Barcelona Supercomputing Center MareNostrum (FI-2023-1-0013); Funded by the European Union. Views and opinions expressed are however those of the author(s) only and do not necessarily reflect those of the European Union, European Commission, European Climate, Infrastructure and Environment Executive Agency (CINEA), or any other granting authority. Neither the European Union nor any granting authority can be held responsible for them (EU Quantum Flagship PASQuanS2.1, 101113690, EU Horizon 2020 FET-OPEN OPTOlogic, Grant No 899794), EU Horizon Europe Program (This project has received funding from the European Union’s Horizon Europe research and innovation program under grant agreement No 101080086 NeQSTGrant Agreement 101080086 — NeQST); ICFO Internal “QuantumGaudi” project; European Union’s Horizon 2020 program under the Marie Sklodowska-Curie grant agreement No 847648; “La Caixa” Junior Leaders fellowships, La Caixa” Foundation (ID 100010434): CF/BQ/PR23/11980043. S. Bandyopadhyay acknowledges AFOSR, USA and Boston University for financial support. 

\appendix
\section{The Fibonacci drive sequence}\label{appA}
The Fibonacci sequence is given by the recursion relation \cite{sutherland1986simple} :
\begin{equation}\label{fibrec}
F_{n+1}=F_n+F_{n-1},
\end{equation}
such that every number is a sum of its previous two numbers. Thus, we have $$F_0=1, F_1=1, F_2=1, F_3=2, F_4=3, F_5=5, F_6=8,$$ and so on. One of the striking features of Fibonacci sequence is that the ratio of two successive Fibonacci numbers quickly approaches the Golden ratio $G$. This can be evaluated as :
\begin{align*}
\frac{X}{Y}&=\frac{X+Y}{X}=G, \\
\frac{X}{Y}&=1+\frac{Y}{X}, \\
G&=1+\frac{1}{G},
\end{align*}
where $X$ and $Y$ are two consecutive Fibonacci numbers. Solving the above equation, we can get the value of $G$ to be $\frac{1+\sqrt{5}}{2}$. From Eq.~\eqref{fibrec}, and the values of $F_0$ and $F_1$, it can be easily seen that the following identity holds true :
\begin{equation}\label{FibId}
F_n=\frac{1}{\sqrt{5}}\left[G^n-\left(-\frac{1}{G} \right)^n \right]
\end{equation}
For large $n$, we can see that $F_n \simeq G^n/\sqrt{5}$ and so we have $F_n/F_{n-1}\approx G$.
\par The drive is done using two distinct square pulse $A$ and $B$, which will follow the Fibonacci sequence. To do this, we shall take the stroboscopic instant to be a Fibonacci number, i.e., $N=F_n$. Within this, we shall have the number of $A$ and $B$ pulses to be equal to $F_{n-1}$ (say, $X$) and $F_{n-2}$ (say, $Y$) respectively. The Fibonacci sequence of pulses is given by 
\begin{equation}\label{fibseq}
ABAABABAABAAB...
\end{equation}
\begin{widetext}
\begin{center}
\begin{tabular}{|c | c | c |} 
 \hline
 Fibonacci sequence & No. of stroboscopic intervals $N$ & No. of $A$'s ($X$) and $B$'s ($Y$) \\ [0.5ex] 
 \hline\hline
 $A$ & 1 ($F_2$) & $X=1 (F_1), Y=0 (F_0)$  \\ 
 $AB$ & 2 ($F_3$) & $X=1(F_2), Y=1(F_1)$ \\
 $ABA$ & 3 ($F_4)$ & $X=2(F_3), Y=1(F_2)$ \\
 $ABAAB$ & 5 ($F_5$) & $X=3(F_4), Y=2(F_3)$  \\
 $ABAABABA$ & 8 ($F_6$) & $X=5(F_5), Y=3(F_4)$ \\
 $ABAABABAABAAB$ & 13 ($F_7$) & $X=8(F_6), Y=5(F_5)$ \\  
 \vdots & \vdots & \vdots\\
 \hline
\end{tabular}
\end{center}
\end{widetext}
Hence for the evolution operator $U(N)$ at a given stroboscopic interval $N=F_n$, we have $X+Y=N$ and $X/Y = G$. Thus, we have 
\begin{equation*}
\frac{X}{Y}+1=\frac{N}{Y}=G+1,
\end{equation*}
\begin{subequations}\label{XY}
\begin{equation}
Y=\frac{N}{1+G},
\end{equation}
\begin{equation}
X=\frac{NG}{1+G}.
\end{equation}
\end{subequations}
Now, to generate the Fibonacci sequence of pulses, we shall define a function
\begin{equation}\label{gen f}
\gamma(l)=\bigl \lfloor (l+1)G\bigr \rfloor-\bigl \lfloor lG\bigr \rfloor,
\end{equation}
where $\lfloor x \rfloor$ denotes the floor function. As we shall see, the function $\gamma(l)$ takes values either 1 or 2, for any positive integer $l$.
\begin{center}
\begin{tabular}{||c|c| c |c |c |c |c |c |c ||}
\hline
$l$ = & 1 & 2 & 3 & 4 & 5 & 6 & 7 & 8 \\
\hline
$\gamma(l)$ = & 2 & 1 & 2 & 2 & 1 & 2 & 1 & 2\\
$(\gamma(l)-1)$ = & 1 & 0 & 1 & 1 & 0 & 1 & 0 & 1\\
$(2-\gamma(l))$ = & 0 & 1 & 0 & 0 & 1 & 0 & 1 & 0\\ 
Sequence & $A$ & $B$ & $A$ & $A$ & $B$ & $A$ & $B$ & $A$\\
\hline
\end{tabular}
\end{center}
We see that the function $(\gamma(l)-1)$ generates a Boolean sequence of 1's and 0's, and if we take 1 corresponding to $A$ pulse, and 0 corresponding to $B$ pulse, we get the exact same sequence as Eq.~\eqref{fibseq}. Thus, at any stroboscopic interval $N$, the sequence can be represented concisely with the function
\begin{equation}\label{seq}
S(N)=[\gamma(N)-1]A + [2-\gamma(N)]B,
\end{equation} 
where $\gamma(N)$ can either take the value 1 (corresponding to $A$ pulse) or the value 2 (corresponding to $B$ pulse). Hence, up to $N$ stroboscopic intervals, the total number of $A$-pulse ($B$-pulse) is given by :
\begin{equation}\label{alp}
\alpha(N)=\sum^N_{l=1}\left[\gamma(l)-1 \right]
\end{equation}
\begin{equation}\label{bet}
\beta(N)=\sum^N_{l=1}\left[2-\gamma(l) \right]=N-\alpha(N)
\end{equation} 

\section{Calculation of evolution operator $U(N)$}\label{appB}
In this appendix, we shall make a perturbative calculation for the unitary (evolution) operator $U(N)$ by explicitly evaluating the BCH expansion. As we had seen in Sec.~\ref{para1}, the unitary operator $U(N)$ on following the Fibonacci sequence ($ABAABABAABAAB...$) involves repeated multiplication of non-commuting matrices $e^A$ and $e^B$, such that $A=-i T H^A_k$ and $B=-i T H^B_k$, hence we have to apply the Baker-Campbell-Hausdorff formula : $e^Ae^B=e^C$, where $C$ is given by :
\begin{equation}
C=A+B+(1/2)[A,B]+(1/12)([A,[A,B]]+[B,[B,A]])+...
\end{equation}
If we evaluate the commutators, we see that order of $T$ associated with it, is equal to the order of the respective commutator, i.e.,
\begin{equation}
[A,B]=-T^2[H^A_k,H^B_k]=-2iT^2\Delta h\sin(k)\sigma^x,
\end{equation}
\begin{align}
[A,[A,B]] &= -iT^3[H^A_k,[H^A_k,H^B_k]]\notag \\
          &= 4i\Delta h T^3\sin(k)\left[(h_A+\cos k)\sigma^y - \sin(k) \sigma^z\right],
\end{align}
where $\Delta h=(h_B-h_A)$. At higher frequencies we can safely ignore the higher order commutators in the BCH expansion. As for lower frequencies too, it will be enough to consider terms upto $\mathcal{O}(T^3)$ to explain the dynamics of the system. Now, we calculate the coefficients of these commutators, by following the Fibonacci sequence. Let $C_n$ be the exponential factor of the unitary operator $U(N)$ at $n$-th stroboscopic interval.
\begin{subequations}
\begin{equation}
U(N=1) = e^A = e^{C_1}
\end{equation}
\begin{equation}\label{c2}
U(N=2)=e^Be^A = e^{C_2}
\end{equation}
\begin{equation}\label{c3}
U(N=3)=e^Ae^Be^A = e^Ae^{C_2}=e^{C_3}
\end{equation}
\begin{equation}\label{c4}
U(N=4)=e^Ae^Ae^Be^A=e^Ae^{C_3}=e^{C_4}
\end{equation}
\begin{equation}\label{c5}
U(N=5)=e^Be^Ae^Ae^Be^A=e^Be^{C_4}=e^{C_5}
\end{equation}
\end{subequations}
Now the exponential factors $C_n$ can be evaluated as (retaining only upto commutators of the order $[A,[A,B]]$ or $[B,[B,A]]$) :\\
\begin{widetext}
\noindent In Eq.~\eqref{c2},
\begin{align}\label{eq c2}
C_2 & = B+A+\frac{1}{2}[B,A]+\frac{1}{12}([A,[A,B]]+[B,[B,A]])\notag \\
 & = \alpha(2)A+\beta(2)B+\delta(2)[A,B]+\eta_1(2)[A,[A,B]]+\eta_2(2)[B,[B,A]]
\end{align}
In Eq.~\eqref{c3},
\begin{align}\label{eq c3}
C_3 & = A+C_2+\frac{1}{2}[A,C_2]+\frac{1}{12}[A,[A,C_2]]+\frac{1}{12}[C_2,[C_2,A]] \notag\\
&= 2A+B-\frac{1}{6}[A,[A,B]]+\frac{1}{6}[B,[B,A]]\notag\\
 & = \alpha(3)A+\beta(3)B+\delta(3)[A,B]+\eta_1(3)[A,[A,B]]+\eta_2(3)[B,[B,A]]
\end{align}
In Eq.~\eqref{c4},
\begin{align}\label{eq c4}
C_4 & = A+C_3+\frac{1}{2}[A,C_3]+\frac{1}{12}[A,[A,C_3]]+\frac{1}{12}[C_3,[C_3,A]]\notag \\
&= 3A+B+\frac{1}{2}[A,B]-\frac{1}{4}[A,[A,B]]+\frac{1}{4}[B,[B,A]]\notag\\
 & = \alpha(4)A+\beta(4)B+\delta(4)[A,B]+\eta_1(4)[A,[A,B]]+\eta_2(4)[B,[B,A]]
\end{align}
In Eq.~\eqref{c5},
\begin{align}\label{eq c5}
C_5 & = A+C_4+\frac{1}{2}[A,C_4]+\frac{1}{12}[A,[A,C_4]]+\frac{1}{12}[C_4,[C_4,A]]\notag \\
&= 3A+2B-[A,B]-\frac{1}{2}[A,[A,B]]\notag\\
 & = \alpha(4)A+\beta(4)B+\delta(4)[A,B]+\eta_1(4)[A,[A,B]]+\eta_2(4)[B,[B,A]]
\end{align}
\end{widetext}
On carefully inspecting the Eqs.~\eqref{eq c2}-~\eqref{eq c5}, it can be shown that if the $N$-th pulse is an $A$-pulse, then the following recursion relation is obeyed:
\begin{subequations}
\begin{equation}\label{A rec1}
\delta(N)=\delta(N-1)+\frac{1}{2}\beta(N-1),
\end{equation}
\begin{multline}\label{A rec2}
\eta_1(N)=\eta_1(N-1)+\frac{1}{2}\delta(N-1)\\+\frac{1}{12}\beta(N-1)(1-\alpha(N-1)),
\end{multline}
\begin{equation}\label{A rec3}
\eta_2(N)=\eta_2(N-1)+\frac{1}{12}\beta^2(N-1).
\end{equation}
\end{subequations}
Similarly, if the $N$-th pulse is a $B$-pulse, then the recursion relations look like :
\begin{subequations}
\begin{equation}\label{B rec1}
\delta(N)=\delta(N-1)-\frac{1}{2}\alpha(N-1),
\end{equation}
\begin{equation}\label{B rec2}
\eta_1(N)=\eta_1(N-1)+\frac{1}{12}\alpha^2(N-,1)
\end{equation}
\begin{multline}\label{B rec3}
\eta_2(N)=\eta_2(N-1)-\frac{1}{2}\delta(N-1)\\+\frac{1}{12}\alpha(N-1)(1-\beta(N-1)).
\end{multline}
\end{subequations}
Now, we shall try to unify the above recursion relations into compact forms by utilising the generating function of Fibonacci sequence, $\gamma(N)$, as defined in Eq.~\eqref{gen f} of Appendix ~\ref{appA}. Similarly, the number of $A$-pulses ($B$-pulses) at the $N$-th stroboscopic interval is given Eq.~\eqref{alp} (Eq.~\eqref{bet}). Thus, the recursion relation for $\delta(N)$ can be written as :
\begin{equation}
\delta(N)=\sum_{m=1}^{N}\left[(\gamma(m)-1)\frac{\beta(m-1)}{2}-(2-\gamma(m))\frac{\alpha(m-1)}{2} \right],
\end{equation}
where $\alpha(0)=\beta(0)=0$. It can be easily shown that if $\gamma(m)=1$, then $\alpha(m-1)=\lfloor mG/(1+G) \rfloor$. Conversely, if $\gamma(m)=2$, then $\beta(m-1)=\lfloor m/(1+G) \rfloor$ (see suppl. material of Ref.~\cite{maity2019fibonacci} for a more direct approach). Substituting these relations, and using the fact $\lfloor mG/(1+G) \rfloor + \lfloor m/(1+G) \rfloor = m-1$, we can simplify the relation for $\delta(N)$ into
\begin{equation}
\delta(N)=\frac{1}{2}\displaystyle\sum_{m=1}^N\left[(\gamma(m)-1)(m-1)-\bigl \lfloor \frac{mG}{1+G}\bigr \rfloor\right].
\end{equation}
Finally, we can use all the above relations to evaluate the coefficients $\eta_1(N)$ and $\eta_2(N)$ as Eq.~\eqref{eta1} and ~\eqref{eta2} of Section ~\ref{sec low freq}, respectively.

\section{Steady state behaviour in the high frequency limit}\label{appC}

In this appendix, we have shown an outline of the steps needed to show that the system reaches a steady state value asymptotically, when we examine a local observable in the high frequency limit (see Ref.~\cite{maity2019fibonacci}). As we had seen in Sec.~\ref{high freq}, for the high drive frequencies , we can retain only upto $\mathcal{O}(T)$ terms in the effective Fibonacci Hamiltonian (Eq.~\ref{eq_ham_first}). Evaluating the commutators explicitly, we can $H_k^{Fib}(N)$ as :
\begin{equation}\label{supp1}
    H_k^{Fib}(N)=a_1 \sigma_x + a_2 \sigma_y + a_3 \sigma_z ,
\end{equation}
where the coefficients $a_i$ are given by\\
\begin{align}
    a_1 &= h_B + \cos k - \frac{\alpha(N)}{N}\Delta h\\
    a_2 &= \sin k\\
    a_3 &= \frac{2\delta(N)}{N}T\Delta h \sin k 
\end{align}
where $\Delta h = h_B - h_A$ is the amplitude difference of the two pulses. Hence, we can see that the effective Fibonacci Hamiltonian is dependent on stroboscopic time N; and so the eigen values and eigen vectors are alos time dependent quantities. From Eq.~\ref{supp1}, we can see that the eigen values of $H_k^{Fib}(N)$ are given by $\epsilon_k^{(1,2)}(N) = \pm \sqrt{a_1^2 + a_2^2 +a_3^3}$, with their corresponding eigenstates as $|\phi_k^{1,2}(N)\rangle$, respectively.\\

The time-dependent Fibonacci Hamiltonian is given by 
\begin{equation}\label{ham}
    H_k(N) = [\gamma (N)-1]H_k^A + [2-\gamma(N)]H_k^B.
\end{equation}
$\gamma(N)$ can be equal to 1 or 2, and thus $H_k(N)$ can be either $H_k^A$ or $H_k^B$ for each $N$. Next, we shall calculate the observable $\varepsilon(N)$ of the system, given by $\varepsilon^{Fib}=(1/L)\displaystyle\sum_{k}\bigl[\langle\psi_k(N)|H_k(N)|\psi_k(N)\rangle - e^g_k(0) ]$. The state after $N$ stroboscopic intervals   can be given as ( in the high frequency approximation ) :
\begin{equation}
    |\psi_k(N)\rangle = e^{-iNTH_k^{Fib}(N)}|\psi_k(N)(0)\rangle,
\end{equation}
where $|\psi_k(N)(0)\rangle$ is the initial state. Working in the basis of eigenstates of $H_k^{Fib}(N)$, we find the $\varepsilon(N)$ of the system in the high frequency limit for a thermodynamically large system with $L\xrightarrow[]{}\infty$ ,
\begin{multline*}
    \varepsilon^{Fib}= \int \frac{dk}{2\pi} \bigl[\langle\psi_k(N)|\underbrace{\sum_{i=1,2}|\phi_k^i(N)\rangle\langle\phi_k^i(N)|}_{\text{=1}}H_k(N)\\ \underbrace{\sum_{j=1,2}|\phi_k^j(N)\rangle\langle\phi_k^j(N)|}_{\text{=1}}\psi_k(N)\rangle - e^g_k(0) \bigr].
\end{multline*}
Rewriting some of the above terms as $c_k^{(1,2)} = \langle \phi_k^{1,2}(N)|\psi_k(0) \rangle $ and matrix elements of $H_k(N)$ in this basis as $H_k^{ij} =  \langle \phi_k^{i}(N)|H_k(N)|\phi_k^j(N) \rangle$, we get 
\begin{multline}\label{eff h}
    \varepsilon^{Fib} = \int \frac{dk}{2\pi} \bigr[|c_k^{(1)}(N)|^2H_k^{11}(N) + |c_k^{(2)}(N)|^2H_k^{22}(N) \\ + (e^{iNT[\epsilon_k^{(1)}(N)-\epsilon_k^{(2)}(N)]}c^{(1)*}_k c^{(2)}_k)H_K^{12}(N)\\ + c.c.)-e^g_k(0) \bigl].
\end{multline}
In the large N limit, i.e. $N\xrightarrow[]{}\infty$, by making use of Riemann-Lebesgue lemma, we can say that the off-diagonal terms having $H^{12}_k(N)$ and its complex conjugate in Eq.~\ref{eff h} oscillate rapidly and hence vanish on integrating over all the $k$-modes. Hence, the steady state expression becomes 
\begin{align}\label{steady}
    \varepsilon^{Fib} &= \int \frac{dk}{2\pi} \bigr[|c_k^{(1)}(N)|^2H_k^{11}(N) + |c_k^{(2)}(N)|^2H_k^{22}(N) - e^g_k(0) \bigl]\nonumber\\
    &= [\gamma(N)-1]\langle H^A \rangle + [2-\gamma(N)]\langle H^B \rangle,
\end{align}
where we have used Eq.~\ref{ham} and the definition of $H_k^{ij}(N)$ and have also replaced the terms as
\begin{equation}\label{steady2}
    \langle H^{A/B} \rangle = \int \frac{dk}{2\pi} \Bigr[ \sum_{i=1,2} |c^i_k(N)|^2 \langle \phi^i_k(N)|H^{A/B}_k|\phi_k^i(N) \rangle \Bigl].
\end{equation}
Numerically, these quantities $\langle H^{A/B} \rangle$ can be obtained using the expectation values of $H^{A/B}$ after $N$ stroboscopic intervals,
\begin{equation}
      \langle H^{A/B}(N) \rangle = \int \frac{dk}{2\pi} \Bigr[ \langle \psi_k(N)|H^{A/B}_k|\psi_k(N) \rangle - e^g_k(0) \Bigl].
\end{equation}
\begin{figure}[h]
	\centering
	
	\includegraphics[width=0.5\textwidth]{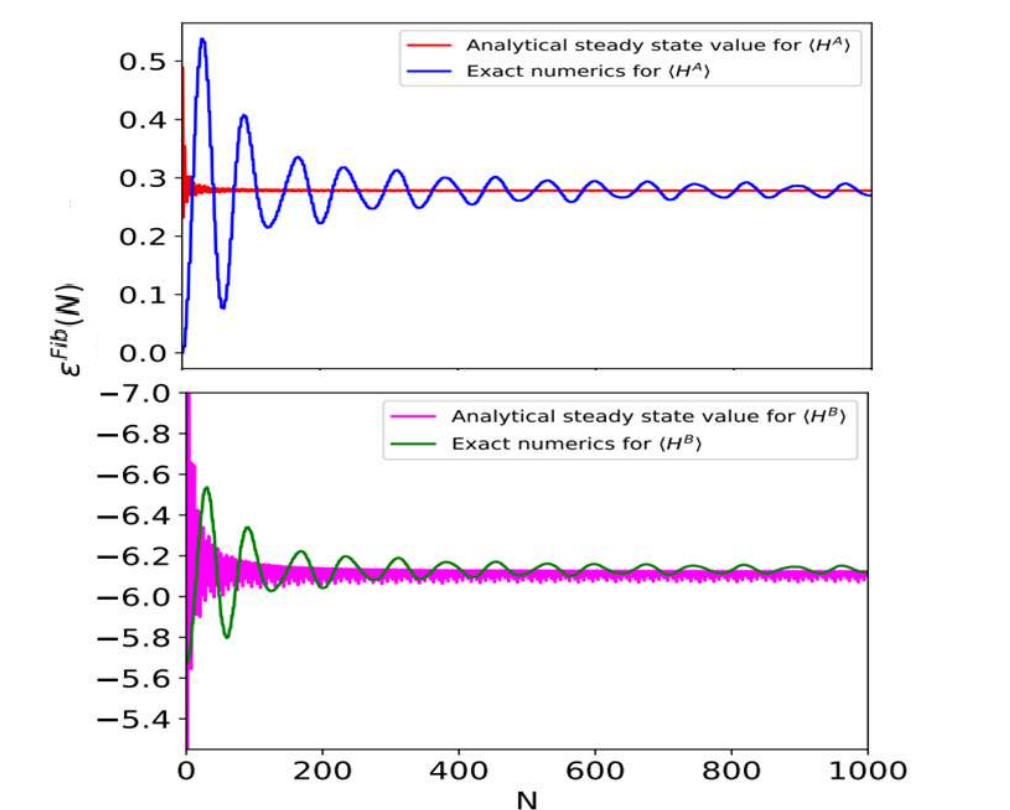}

	\caption{$\langle H^A \rangle$ and $\langle H^B \rangle$ plotted as function of $N$ for a Fibonacci drive protocol with $h_A = 1$, $h_B = 10$, $\omega=500$. The solid blue (green) lines show the results obtained numerically for $\langle H^A \rangle (\langle H^B\rangle)$, and the red (magenta) lines are the corresponding steady state values using the analytical formula obtained in Appendix~\ref{appC} (Eq.~\ref{steady2}). The total value of $\varepsilon_O(N)$ of the system oscillates between these two steady state values as per the Fibonacci sequence as shown in Eq.~\ref{steady}.}\label{steady_fig}
\end{figure}

From Fig.~\ref{steady_fig}, we can see that in the limit of large $N$, the observable $\varepsilon(N)$ rapidly converges to the analytical steady state values $\langle H^{A/B} \rangle$, given by Eq.~\ref{steady2}. This is because the contribution of the off-diagonal terms vanishes on summation over all the $k$-modes, due to the rapidly fluctuating $\delta(N)/N$ term. The steady state value is independent of time, and is determined by the behaviour of $\alpha(N)/N$. From Eq.~\ref{steady}, we can see that the $\varepsilon(N)$ of the total Hamiltonian varies between these two steady state values following the Fibonacci sequence. 

\begin{figure}[t]
	\centering
	
		\includegraphics[width=0.45\textwidth]{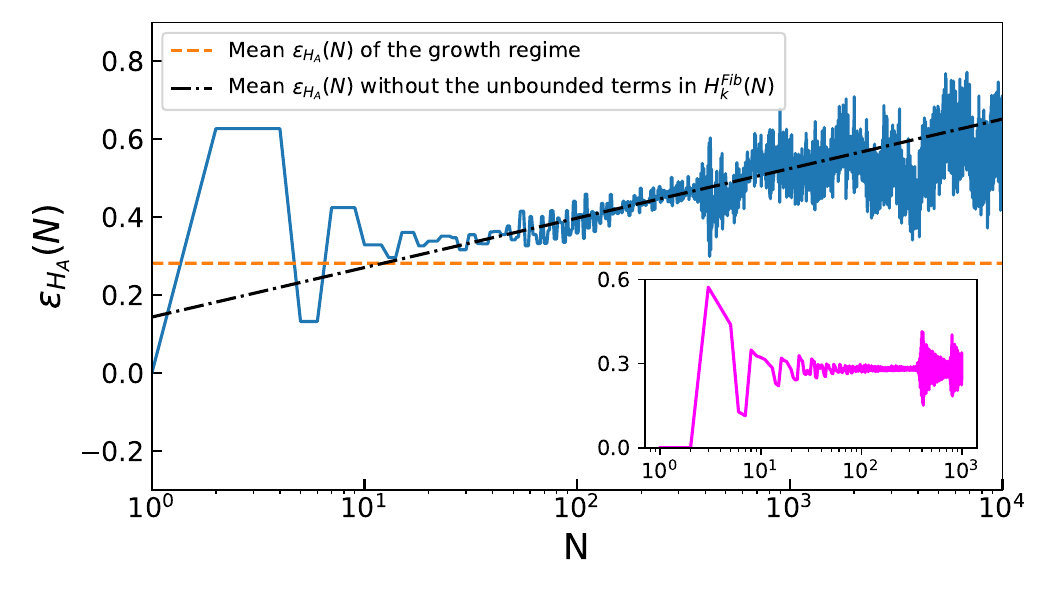}
		\label{fitting}
	\caption{Numerical method of finding out the crossover time $N_{c}$ using the method described in Appendix ~\ref{appC}. Here, the observable $\varepsilon_{H_A}(N)$ is plotted as a function of $N$ for $\omega=50, L=100$, initial state to be the ground state of Hamiltonian $H_A$, both with and without the unbounded terms in the effective Hamiltonian $H_k^{Fib}(N)$ and their point of intersection is taken to be $N_{c}$; the inset shows that $\varepsilon_{H_A}(N)$ using only the bounded terms in the effective Hamiltonian (Eq.~\eqref{eq_ham_first}) is constant with time, in contrast to the heating behaviour obtained using exact numerical diagonalization.}
\end{figure}

\section{Numerical evaluation of the crossover time $N_{c}$ at any given frequency }\label{appD}
In this appendix, we shall see the technique used to numerically evaluate the crossover time, $N_{c}$ from the pre-thermal regime to the growth regime. To make a rough estimate of the same, we shall employ the following method : i) Plot the local observable $\varepsilon_{H_A}(N)$ using only the bounded terms in the effective Hamiltonian, i.e., Eq.~\eqref{eq_ham_first}. The mean value of $\varepsilon_{H_A}(N)$ will be constant with time, as shown in inset of Fig.~\ref{fitting}. ii) Fit a straight line into the growth regime of $\varepsilon_{H_A}(N)$ obtained by exact diagonalization of the system, i.e., using the exact relation for unitary operator $U(N)$ without omitting any term. iii) Finding out the point of intersection of these two straight lines, which will be $N_{c}$ (as shown in Fig. \ref{fitting}).\\
It should be kept in mind that the crossover from pre-thermal to the heating regime is not instantaneous, instead it is spread out over a few stroboscopic instants, and hence, the crossover time obtained is nothing but a rough estimate. We have taken the most consistent set of data for our analysis after repeated application of the above procedure.

\begin{figure}[t]
	\centering
	
		\includegraphics[width=0.45\textwidth]{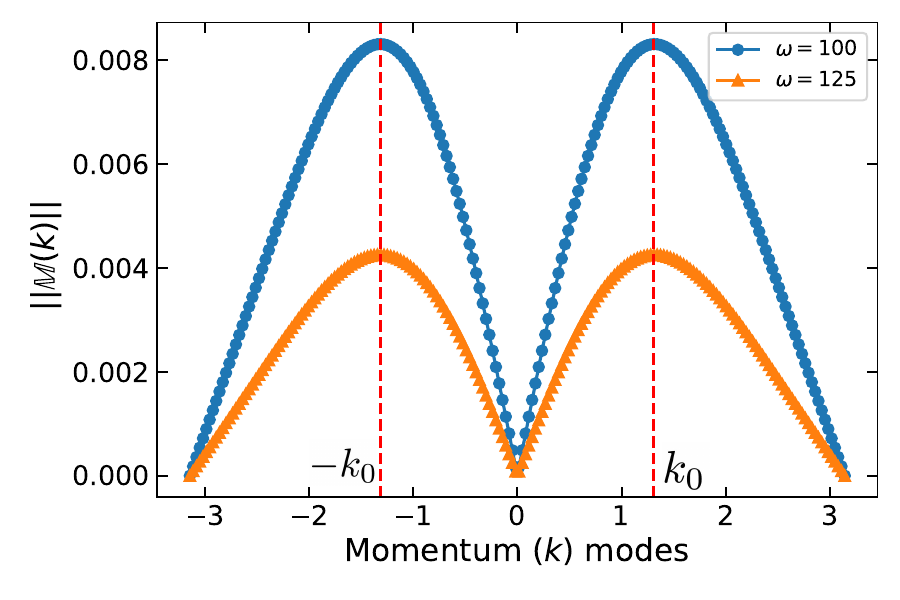}
		
	\caption{Scaling of the average norm of the coefficients of the order $\mathcal{O}(T^3)$ of the effective Hamiltonian $H_k^{Fib}(N)$ with the momentum $k$ modes for two different frequencies. We see that the maximum contribution to the average norm comes from two $k$-modes: $k_0$ and $-k_0$, and the positions of the maxima are same for all frequencies. For both of the above analyses, we have used the same initial conditions as before : the initial state taken to be the ground state of Hamiltonian $H_A$; $h_A=1$, $h_B=10$ and system size $L=100$.}\label{k spacing1}
\end{figure}

\section{Dependence of prethermal lifetime on commutator norm of the effective Hamiltonian}\label{appE}
In this subsection, we discuss the dependence of the prethermal regime on the commutator norm of the highest order unbounded term in the effective Hamiltonian. This can be seen from the scaling of the average norm of the coefficients of the order $\mathcal{O}(T^3)$ of the effective Hamiltonian $H_k^{Fib}(N)$ with the momentum $k$ and the analysis given below.
\par As explained in Sec.~\ref{drive freq}, the system starts absorbing energy when terms of order $\mathcal{O}(T^3)$ become dominant. When observed at Fibonacci instants $F(\mathcal{N})$, the effective Hamiltonian for a single $k$-mode can be written as a sum of the bounded terms and the unbounded terms :
\begin{equation}
H_k^{Fib}(F(\mathcal{N}))= H_k^{bounded}(F(\mathcal{N}))+H_k^{unbounded}(F(\mathcal{N}))
\end{equation}
The leading order unbounded term of $H_k^{unbounded}(F(\mathcal{N}))$ is of the form :
\begin{equation}
H_k^{unbounded}(F(\mathcal{N}))\approx G^{\mathcal{N}}T^3 \lVert \mathbb{M}(k) \rVert
\end{equation}
where we have used the fact that the NECs $\mu_i(F(\mathcal{N}))/F(\mathcal{N})$, where $\{i=1,2,3\}$, vary as $G^{\mathcal{N}}$ when observed at Fibonacci instants (see Eqs. ~\eqref{mu1}-~\eqref{mu3}). In the above equation, $\lVert \mathbb{M}(k) \rVert$ is the average norm of  $[H_k^A,[H_k^A,[H_k^A,H_k^B]]]$, $[H_k^B,[H_k^B,[H_k^A,H_k^B]]]$ and $[H_k^A,[H_k^B,[H_k^A,H_k^B]]]$. Note that in calculating the above, we have once again made use of the fact that the stroboscopic instant $N$ corresponding to Fibonacci instant $\mathcal{N}$ is given as $N\approx G^{\mathcal{N}}$ (see Appendix ~\ref{appA}). On summing over all the $k$-modes, the leading order unbounded term becomes :
\begin{align}\label{Mk}
\langle H^{unbounded} \rangle &\approx \sum_k H^{unbounded}_k~~, \notag\\
\implies \langle H^{unbounded} \rangle &\sim N_{c}T^3\sum_k \lVert \mathbb{M}(k) \rVert
\end{align}
It can be seen from Fig.~\ref{k spacing1} that the average norm $\lVert \mathbb{M}(k) \rVert$ has two maxima at $k_0$ and $-k_0$. This ensures that the maximum contribution to the norm in the leading order unbounded term of Eq.~\eqref{Mk} comes from these two $k$-modes. This gives a more rigorous estimation of the order at which the effective Hamiltonian expansion breaks down, which in turn results in the onset of heating.

\bibliographystyle{ieeetr}
\bibliography{citation}
\end{document}